\documentclass[reprint,nofootinbib]{revtex4}
\usepackage{graphicx}  
\usepackage{dcolumn}   
\usepackage{bm}        
\usepackage{amssymb}
\usepackage{hyperref}
\usepackage{multirow}
\usepackage{amsmath}
\usepackage{cases}
\usepackage{placeins}
\usepackage{subfig}
\usepackage{xcolor}

\begin{document}
\title{Consequences of $b\, \to\, s\, \mu^+\, \mu^-$ anomalies on $B \, \to \, K^{(*)} \,\nu \, \bar{\nu}$, 
$B_s \to\, (\eta,\eta') \, \nu\,\bar{\nu}$ and $B_s \, \to \, \phi \, \nu \, \bar{\nu}$ decay observables}
 \author{N~Rajeev${}$}
 \email{rajeev\_rs@phy.nits.ac.in} 
 \author{Rupak~Dutta${}$}
 \email{rupak@phy.nits.ac.in}
 \affiliation{${}$ National Institute of Technology Silchar, Silchar 788010, India}

\begin{abstract}
 
The long persistent discrepancies in $b\,\to\,s\, \ell^+\, \ell^-$ quark level transitions continue to be the ideal platform for an indirect 
search of beyond the standard model physics. The recent updated measurements of $R_K$, $\mathcal{B}(B_s\, \to\, \phi\, \mu^+\,\mu^-)$ 
and $\mathcal{B}(B_s\, \to\, \mu^+\,\mu^-)$ from LHCb deviate from the standard model expectations at more than $3\sigma$ level. 
Similarly, measurements of $R_{K^*}$ and $P_5^{\prime}$ in $B \to K^{*} \, \ell^+\,\ell^-$ decays
disagree with the standard model predictions at $\sim 2.4\sigma$ and $\sim 3.3\sigma$, respectively. 
Moreover, recent measurement of ratio of branching ratios $R_{K_S^0}$ and $R_{K^{*+}}$ in $B^0 \, \to \, K_S^{0} \, \ell^+\,\ell^-$ and 
$B^+ \, \to \, K^{+*} \, \ell^+\,\ell^-$ decays deviate from the standard model prediction at $1.4\sigma$ and $1.5\sigma$, respectively. 
Considering the $R_{K_S^0} - R_{K^{*+}}$ combination, the difference
with the SM predictions currently stands at about $2\sigma$.
Motivated by these anomalies we search for the patterns of new physics 
in the family of flavor changing neutral current decays with neutral leptons in the final state undergoing $b \, \to \,s\, \nu \, \bar{\nu}$ 
quark level transitions.
There are close relations between $b\,\to\,s\, \ell^+\, \ell^-$ and $b \, \to \,s\, \nu \, \bar{\nu}$ transition decays not only in the 
standard model but also in beyond the standard model physics. In beyond the standard model physics the left handed charged leptons can be
related to the neutral leptons via $\rm SU(2)_L$ gauge symmetry. Moreover, there are several advantages of studying 
$b \, \to \,s\, \nu \, \bar{\nu}$ transitions over $b\,\to\,s\, \ell^+\, \ell^-$ as they are
free from various hadronic uncertainties such as the non-factorizable corrections and photonic penguin contributions.
In this context, we use the standard model effective field theory formalism and explore the consequences of $b\, \to\, s\, \mu^+\, \mu^-$ 
anomalies on $B \, \to \, K^{(*)} \,\nu \, \bar{\nu}$, $B_s \to\, (\eta,\eta') \, \nu\,\bar{\nu}$ and 
$B_s \, \to \, \phi \, \nu \, \bar{\nu}$ decay observables in several 1D and 2D new physics scenarios.

\end{abstract}

\maketitle

\section{Introduction}
In the standard model~(SM), the three families of leptons are identical except for their masses. More specifically, the photon, $Z$ and 
$W^{\pm}$ bosons 
couple to them with equal strengths. Hence the SM is lepton flavor universal. However, there exists several hints of lepton flavor
universality~(LFU) violation in 
$B$ meson decays mediating via $b \to c\,\ell\,\nu$ charged current and $b \to s\,\ell^+\,\ell^-$ neutral current transitions reported by BABAR, 
Belle, and more recently by LHCb Collaboration.
The recent updated measurements of $R_K$, $\mathcal{B}(B_s\, \to\, \phi\, \mu^+\,\mu^-)$, $\mathcal{B}(B_s\, \to\, \mu^+\,\mu^-)$ and  
new measurements of $R_{K^0_S}$ and $R_{K^{*+}}$ from LHCb
continue to exhibit the same pattern of deviations with respect to the SM expectations.
For completeness, we 
report the current status of several $b\,\to\,s\,\ell^+\,\ell^-$ decay observables in Table~\ref{status}.

The rare semileptonic $b\,\to\,s\,\ell^+\,\ell^-$ transition processes are very interesting probes of new physics~(NP) because of their 
sensitivity to various NP contributions that can, in principle, appear in the penguin loop diagrams or in the box diagrams. It is, however, 
worth
mentioning that a precise SM prediction of the observables is crucial to disentangle the genuine NP contribution from the SM
uncertainties that may come from meson to meson transition form factors and CKM matrix elements. In recent years, the QCD motivated 
approaches based on the lattice quantum chromodynamics (LQCD) and light cone sum rule (LCSR)
have provided very precise value of the form factors for various $b\,\to\,s\,\ell^+\,\ell^-$ processes. 
For several preferred decay modes such as $B \, \to \, K^{(*)} \, \ell^+\,\ell^-$ and $B_s \, \to \, \phi \, \ell^+\,\ell^-$, a very precise
value of the form factors are obtained within LCSR and LQCD~\cite{Bouchard:2013eph,Bharucha:2015bzk} approach.
Apart from these hadronic uncertainties, there exists several other challenges such as short distance contributions, non-local effects below 
the charmonium, the hadronic non-local effects, the non-factorizable effects arising due to spectator scattering and finite width 
effects~\cite{Khodjamirian:2010vf,
Khodjamirian:2012rm,Bobeth:2017vxj,Gubernari:2020eft,Cheng:2017smj,Descotes-Genon:2019bud} that can cause $b\,\to\,s\,\ell^+\,\ell^-$ neutral 
processes more difficult to access theoretically.  
Although these corrections tend to increase the discrepancy in the branching fractions, the normalized angular observables such as $P'_5$ and 
other LFU sensitive ratios are mostly insensitive to these corrections. 
Moreover, a global fit including all these corrections are still awaited~\cite{Descotes-Genon:2019bud,Virto:2021pmw}.  

\begin{table}[ht]
\centering
\setlength{\tabcolsep}{6pt} 
\renewcommand{\arraystretch}{1.5} 
\resizebox{\columnwidth}{!}{
\begin{tabular}{|c|c|c|c|c|}
\hline
       & $q^2$ bins & Theoretical predictions & Experimental measurements & Deviation\\
\hline
\hline
$R_K$  & [1.1, 6.0] & $1 \pm 0.01$~\cite{Bordone:2016gaq,Hiller:2003js} & $0.846^{+0.044}_{-0.041}$~\cite{LHCb:2021trn,LHCb:2019hip} & $\sim3.1\sigma$ \\
\hline
$R_{K_S^0}$  & [1.1, 6.0] & $1 \pm 0.01$ & $0.66^{+0.20}_{-0.14}$ (stat) $^{+0.02}_{-0.04}$ (syst)~\cite{LHCb:2021lvy} & $\sim 1.4\sigma$ \\
\hline
\multirow{4}{*}{$R_{K^*}$} & \multirow{2}{*}{[0.045, 1.1]} & $1 \pm 0.01$~\cite{Bordone:2016gaq,Hiller:2003js} & $0.660^{+0.110}_{-0.070}$ (stat) $\pm 0.024$ (syst)~\cite{Aaij:2017vbb} & \multirow{4}{*}{$\sim 2.2-2.5\sigma$}\\ 
                                                        &  & $1 \pm 0.01$~\cite{Bordone:2016gaq,Hiller:2003js} & $0.52^{+0.36}_{-0.26}$ (stat) $\pm 0.05$ (syst)~\cite{Abdesselam:2019wac} & \\ \cline{2-4}
                           & \multirow{2}{*}{[1.1, 6.0]}   & $1 \pm 0.01$~\cite{Bordone:2016gaq,Hiller:2003js} & $0.685^{+0.113}_{-0.069}$ (stat) $\pm 0.047$ (syst)~\cite{Aaij:2017vbb} & \\
                                                       &   & $1 \pm 0.01$~\cite{Bordone:2016gaq,Hiller:2003js} & $0.96^{+0.45}_{-0.29}$ (stat) $\pm 0.11$ (syst)~\cite{Abdesselam:2019wac} & \\
\hline
$R_{K^{*+}}$  & [0.045, 6.0] & $1 \pm 0.01$ & $0.70^{+0.18}_{-0.13}$ (stat) $^{+0.03}_{-0.04}$ (syst)~\cite{LHCb:2021lvy} & $\sim 1.5\sigma$ \\
\hline
\multirow{3}{*}{$P_{5}^{\prime}$} & \multirow{1}{*}{[4.0, 6.0]} & $-0.757 \pm 0.074$~\cite{Descotes-Genon:2013vna} & $-0.21 \pm 0.15$~\cite{Aaboud:2018krd,Aaij:2013qta,Aaij:2015oid} & $\sim3.3\sigma$\\ \cline{2-5}
                                  & \multirow{1}{*}{[4.3, 6.0]} & $-0.774^{+0.0.061+0.087}_{-0.059-0.093}$~\cite{DescotesGenon:2012zf} & $-0.96^{+0.22}_{-0.21}$ (stat) $\pm 0.16$ (syst)~\cite{CMS Collaboration} & $\sim1.0\sigma$\\  \cline{2-5}
                                  & \multirow{1}{*}{[4.0, 8.0]} & $-0.881 \pm 0.082$~\cite{Descotes-Genon:2014uoa} & $-0.267^{+0.275}_{-0.269}$ (stat) $\pm 0.049$ (syst)~\cite{Abdesselam:2016llu} & $\sim2.1\sigma$\\
\hline
$\mathcal{B}(B_s\, \to\, \phi\, \mu^+\, \mu^-)$ & {[1.1, 6.0]} & {$(5.37 \pm 0.66)\times 10^{-8}$}~\cite{Aebischer:2018iyb,Straub:2015ica} & {$(2.88 \pm 0.22)\times 10^{-8}$}~\cite{LHCb:2021zwz,LHCb:2013tgx,LHCb:2015wdu} & $\sim3.6\sigma$\\
\hline
$\mathcal{B}(B_s\, \to \mu^+\, \mu^-)$ & - & {$(3.66 \pm 0.14)\times 10^{-9}$}\cite{Bobeth:2013uxa,Beneke:2019slt} & {$(3.09^{+0.46+0.15}_{-0.43-0.11})\times 10^{-9}$}~\cite{LHCb:2021vsc} & - \\
\hline
\hline
\multirow{2}{*}{$\mathcal{B}(B^+ \to K^+ \nu \nu)$} & - & \multirow{2}{*}{$(4.4 \pm 0.07)\times 10^{-6}$~\cite{zenodo.5543714}} & $< 1.6 \times 10^{-5}$~\cite{BaBar:2013npw} & - \\
                                                    &   &                                                                       & $< 4.1 \times 10^{-5}$~\cite{Dattola:2021cmw} &  \\
\hline
$\mathcal{B}(B^0 \to K^0 \nu \nu)$ & - & $(4.1 \pm 0.05)\times 10^{-6}$~\cite{zenodo.5543714} & $< 2.6 \times 10^{-5}$~\cite{Belle:2017oht} & - \\
\hline
$\mathcal{B}(B^0 \to K^{0*} \nu \nu)$ & - & $(9.5 \pm 0.09)\times 10^{-6}$~\cite{zenodo.5543714} & $< 1.8 \times 10^{-5}$~\cite{Belle:2017oht} & - \\
\hline
$\mathcal{B}(B^+ \to K^{+*} \nu \nu)$ & - & $(10 \pm 1)\times 10^{-6}$~\cite{zenodo.5543714} & $< 4.0 \times 10^{-5}$~\cite{Belle:2013tnz} & - \\
\hline
\end{tabular}}
\caption{Current status of $b\,\to\,s\,\ell^+\,\ell^-$ on $b \, \to \,s\, \nu \, \bar{\nu}$ decay observables}
\label{status}
\end{table}

Similar to the family of neutral decays with charged leptons in the final state undergoing $b\, \to\, s\, \ell^+\, \ell^-$ quark level 
transition, there also exist another family of flavor changing neutral transitions with two neutral leptons in the final state. 
Study of rare processes mediating via $b \, \to \,s\, \nu \, \bar{\nu}$ quark level transitions are important for several reasons.
First, these processes are theoretically cleaner than the corresponding neutral current 
decays with two charged leptons in the final state as they do not suffer from hadronic uncertainties
beyond the form factors such as the non-factorizable corrections and photon penguin contributions. Second, 
the $b\,\to\,s\, \ell^+\, \ell^-$ and $b \, \to \,s\, \nu \, \bar{\nu}$ transition decays are very closely related not only in the
SM but also in beyond the standard model physics. 
Hence, study of these decay modes theoretically as well as experimentally will be crucial to look for potential new physics proposed to 
explain the anomalies present in $b\,\to\,s\,\ell^+\,\ell^-$ transition decays.

Study of $b \, \to \,s\, \nu \, \bar{\nu}$ transition decays are experimentally challenging because of the presence of the neutral leptons 
which leave no information in the detectors. There exist a few experiments that predict the upper bound of the branching ratio (BR) of 
$B \, \to \, K^{(*)} \,\nu \, \bar{\nu}$ decays. 
The initial experimental study on $b \, \to \,s\, \nu \, \bar{\nu}$ channels was done by BaBar~\cite{BaBar:2004xlo} in 2004, 
where the upper limit of $\mathcal B(B^+ \, \to \, K^{+} \,\nu \, \bar{\nu})$ and $\mathcal B(B^+ \, \to \, \pi^{+} \,\nu \, \bar{\nu})$ at 
90\% CL were reported 
using the hadronic reconstruction method. Later the results were updated in 2008~\cite{BaBar:2008wiw} and in 2013~\cite{BaBar:2013npw},
respectively. Similarly, the first result by Belle~\cite{Belle:2007vmd} was published in the year 2007. Subsequently the results were 
updated in 2013 and 2017. So far all the measurements used tagged approaches where the second $B$ meson that is produced in 
$e^+ e^- \to \varUpsilon (4S) \to B \bar{B}$ was explicitly reconstructed either in a hadronic decay or in 
semileptonic decay~\cite{BaBar:2013npw,Belle:2013tnz,Belle:2017oht}. 
This approach of tagging, in principle, suppresses the background events and results in a low signal reconstruction efficiency which is 
typically below 1\%.
Very recently, Belle II used a novel technique based on the inclusive tagging methods and exploited the topological features of 
$B^+\, \to K^+\, \nu\, \bar{\nu}$ decays.
This inclusive tagging method has helped to identify $B^+\, \to K^+\, \nu\, \bar{\nu}$ from seven dominant background processes of the 
generic B mesons decays. It improves the signal efficiency by 4\% at the cost of higher background levels in comparison to earlier methods. 
An upper bound of $\mathcal B(B^+\, \to K^+\, \nu\, \bar{\nu}) < 4.1 \times 10^{-5}$~\cite{Dattola:2021cmw} at 90\% CL
is reported very recently. Combining this with earlier measurements from Belle and BaBar, the estimated world average
for $\mathcal B(B^+\, \to K^+\, \nu\, \bar{\nu})$ is reported to be $(1.1\pm 0.4)\times 10^{-5}$~\cite{Dattola:2021cmw}. We summarize all the 
results in Table~\ref{status}.

Our main aim is to explore the consequences of $b\,\to\,s\, \ell^+\, \ell^-$ anomalies on several $b \, \to \,s\, \nu \, \bar{\nu}$ transition
decays in a model independent effective theory formalism. Theoretical study on $b \, \to \,s\, \nu \, \bar{\nu}$ transition decays are 
limited as compared to $b\,\to\,s\,\ell^+\,\ell^-$ transition decays~\cite{Altmannshofer:2021qrr,Hurth:2020ehu,Descotes-Genon:2012isb,
Aebischer:2018iyb,Capdevila:2017bsm,Rajeev:2020aut,Alok:2010zd,Dutta:2019wxo,Alguero:2021anc,Geng:2021nhg,Isidori:2020acz,Datta:2019zca,
Alguero:2019pjc,Altmannshofer:2017poe,BhupalDev:2021ipu,
Altmannshofer:2020axr,MunirBhutta:2020ber,Carvunis:2021jga,Alok:2019ufo} and to the $b\,\to\,c\,\ell\,\nu$ decays~\cite{Bifani:2018zmi,
Dutta:2013qaa,Azatov:2018knx,Dutta:2018vgu,Alok:2017qsi,Jung:2018lfu,Dutta:2018jxz,Murgui:2019czp,Rajeev:2018txm,
Rajeev:2019ktp,Das:2019cpt,Das:2021lws,Dutta:2017xmj}.
It is well known that $b\,\to\,s\, \ell^+\, \ell^-$ and $b \, \to \,s\, \nu \, \bar{\nu}$ transition decays are related not only in SM but 
also in beyond the standard model physics. In beyond the standard model physics, they are related via $\rm SU(2)_L$ gauge symmetry and can be
best exploited using SM effective field theory~(SMEFT) formalism. 
The concept of $\rm SU(2)_L$ gauge symmetry was established earlier in few literatures~\cite{Bhattacharya:2014wla,Bhattacharya:2016mcc,
Calibbi:2015kma,Alonso:2015sja,Hiller:2014yaa,Glashow:2014iga,Altmannshofer:2014rta,Ghosh:2014awa,Hiller:2014ula} to
provide a simultaneous explanation of $b\, \to c\, \ell\, \nu$ and $b\,\to\,s\,\ell^+\,\ell^-$ anomalies.
The authors in Ref.~\cite{Bhattacharya:2014wla,Bhattacharya:2016mcc} point out that by assuming the new physics scale much larger than 
the weak scale, the operators can be made invariant under $SU(3)_C \times SU(2)_L \times U(1)_Y$ gauge group. 
There arises two consequences. First, the left-handed fermion fields must be replaced by $\rm SU(2)_L$ doublet and second, there will be 
two new physics operators that are invariant under $\rm SU(2)_L$.
As a result, these new physics operators lead to different type of contributions to the neutral current and the charged current interactions,
which, in turn, can be used to explain $R_K$ and $R_{D^{(*)}}$ anomalies simultaneously.
We list out few more relevant literatures on $b \, \to \,s\, \nu \, \bar{\nu}$ transition decays~\cite{Altmannshofer:2009ma,
Buras:2014fpa,Descotes-Genon:2021doz,Bause:2021ply,Browder:2021hbl,Kahn:2019abn,Maji:2018gvz,Ahmady:2018fvo,Fajfer:2018bfj,Bordone:2017lsy,
Das:2017ebx,Niehoff:2015qda,Sahoo:2015fla,Buras:2015yca,Calibbi:2015kma,Girrbach-Noe:2014kea,Mohapatra:2021ynn,Felkl:2021uxi,
Biancofiore:2014uba,Colangelo:1996ay,Buchalla:1998ba,Bartsch:2009qp,He:2021yoz,Alda:2021rgt} where such connections have been addressed.
More specifically, in Ref.~\cite{Altmannshofer:2009ma}, the authors did study $B \, \to \, K^{(*)} \,\nu \, \bar{\nu}$ decays  
in the SM as well as in several NP models such as MSSM, modified $Z/Z'$ penguins, single scalar extension.
The authors also pointed out the correlations of $K\to \pi \nu \, \bar{\nu}$,
$K_L \to \pi \nu \, \bar{\nu}$, $B \to X_s \ell \ell$ and $B_s \to \mu \mu$ in case of right handed NP.
Similarly, in Ref.~\cite{Buras:2014fpa}, the authors study $B \, \to \, K^{(*)} \,\nu \, \bar{\nu}$ decays in the SM and in several beyond the
SM models such as $Z'$ model, MSSM, leptoquark model. They also use the model independent SMEFT framework and explored several NP scenarios.
Very recently, in Ref.~\cite{Descotes-Genon:2021doz}, the authors study the implication of $b\,\to\,s\,\ell\,\ell$ anomalies on several
$b \, \to \,s\, \nu \, \bar{\nu}$ and $s \, \to \,d\, \nu \, \bar{\nu}$ decays.
They also discussed the correlation between $B \, \to \, K^{(*)} \,\nu \, \bar{\nu}$ and $K\to \pi \nu \, \bar{\nu}$ decays in the case of
minimal flavor violation.
In Ref.~\cite{Bause:2021ply}, the authors used the SMEFT framework and
estimated the new limit on the branching ratios of $B \to \, (K,X_s) \,\nu \, \bar{\nu}$, $B_s \, \to \, \phi \, \nu \, \bar{\nu}$
and $B \to \, (\pi,\rho) \,\nu \, \bar{\nu}$ decays. 
In Ref.~\cite{Browder:2021hbl}, the authors have explored the possibility of enhancement in the branching ratio of 
$B \, \to \, K \,\nu \, \bar{\nu}$ using NP within scalar and vector leptoquarks and generic vector gauge boson $Z'$ model assuming minimal
new particle content. In the Ref.~\cite{He:2021yoz} the authors investigate $B \, \to \, K^{(*)} \,\nu \, \bar{\nu}$ decays in the context of 
non-standard neutrino interactions. Moreover, in Ref.~\cite{Alda:2021rgt}, the authors use the SMEFT framework and perform a global fit to 
the $R_{D^{(*)}}$ and $R_{K^{(*)}}$ data. They, indeed, find a strong correlation between the $C_L$ operator of 
$b \, \to \,s\, \nu \, \bar{\nu}$ and $C_{V_L}$ of $b\,\to\,c\,\ell\,\nu$ decays. 

In the present article, we study the implication of $b\,\to\,s\, \ell^+\, \ell^-$ anomalies on $B \, \to \, K^{(*)} \,\nu \, \bar{\nu}$, 
$B_s \to\, (\eta,\eta') \, \nu\,\bar{\nu}$ and $B_s \, \to \, \phi \, \nu \, \bar{\nu}$ decay observables
within the SMEFT framework. We give predictions of the branching fractions and longitudinal polarization fraction in the SM as well as in the 
presence of several 1D and 2D new physics scenarios constructed from SMEFT operators. We perform a global fit to the 
$b\,\to\,s\, \ell^+\, \ell^-$ data to obtain the allowed new physics parameter space. Our fit analysis include the experimental measurements of
$R_K$, $R_{K^*}$, $P_5^{\prime}$, $\mathcal{B}(B_s\, \to\, \phi\, \mu^+\,\mu^-)$ and $\mathcal{B}(B_s\, \to\, \mu^+\,\mu^-)$ and, 
in particular, we make use of the latest updated measurements of $R_K$, $\mathcal{B}(B_s\, \to\, \phi\, \mu^+\,\mu^-)$ and $\mathcal{B}(B_s\, 
\to\, \mu^+\,\mu^-)$. In addition, we also check the compatibility of the constrained new physics parameter space with the 
latest $b \, \to \,s\, \nu \, \bar{\nu}$ experimental data. 

So far we don’t have many experimental results on $b \, \to \,s\, \nu \, \bar{\nu}$ transition decays. The experimental techniques used for
$B \, \to \, K^{(*)} \,\nu \, \bar{\nu}$ can be used for $B_{s} \to \, (\eta^{(\prime)},\phi) \,\nu \, \bar{\nu}$ decays as well.
Currently, Belle II can be the ideal platform to perform such analysis and predict the upper bound of the branching fractions of these decays.
In contrast to the $\varUpsilon (4S)$ resonance at Belle where it goes to $B\bar{B}$ pair, Belle II runs at $\varUpsilon (5S)$ as well.
The $\varUpsilon (5S)$ goes into pairs of $B$ or $B_s$. 
Belle II has collected samples at the $\varUpsilon (5S)$ resonance at an integrated luminosity of $121.4\, \rm fb^{-1}$. By taking the 
cross-section
for $e^- e^+ \to b\bar{b}$ and $\mathcal{B}(\varUpsilon (5S)) \to B_s^{(*)0}\bar{B}_s^{(*)0}=0.172 \pm 0.030$~\cite{Belle:2012tsw}, one can
estimate a total $(7.11 \pm 1.30)\times 10^{6}$ $B_s^{(*)0}\bar{B}_s^{(*)0}$ pairs at the KEKB collider. Since only a fraction of these 
$B_s$ decays will survive the kinematics, we expect the statistical uncertainty for $B_{s} \to \, (\eta^{(\prime)},\phi) \,\nu \, \bar{\nu}$
to be more with respect to 
$B \, \to \, K^{(*)} \,\nu \, \bar{\nu}$ decay channel. Moreover, the missing momentum in the final state due to undetected neutrinos can 
cause difficulties in reconstructing these channels.

The paper is organized as follows. In Sec.~\ref{pheno}, we start with a brief overview of the standard model effective field theory and 
write down the effective Hamiltonian governing $b \to s\, \nu\, \bar{\nu}$ and $b\,\to\,s\,\ell^+\,\ell^-$ decays. 
In Sec.~\ref{result}, we give predictions of all the
observables in the SM and in several 1D and 2D NP scenarios. We conclude with a brief summary of
our results in Sec.~\ref{concl}.

\section{Phenomenology}\label{pheno}
\subsection{Standard model effective field theory}
So far LHC searches do not provide any direct evidence of new particles close to the electroweak scale. It indirectly suggests the 
existence of NP at a scale that must lie beyond the electroweak scale. A better way to look for indirect signature of NP in a model 
independent basis can be attained by considering SM effective field theory~(SMEFT) framework. The SMEFT Lagrangian contains all possible set 
of higher dimensional operators that are built out of the SM fields and are consistent with the $SU(3)_c \times SU(2)_L \times U(1)_Y$ gauge 
group. In SMEFT, the higher dimensional operators are suppressed by appropriate power of the NP scale. For a complete set of dimension
six and dimension eight operators, we refer to Refs.~\cite{Buchmuller:1985jz,Arzt:1994gp,Grzadkowski:2010es,Murphy:2020rsh,Li:2020gnx}. 
It is well known that the left handed charged leptons are related to the neutral leptons via $SU(2)_L$ symmetry.
In this context, the SMEFT framework can be a powerful tool to study the correlation between $b \to s\,\ell^+\ \ell^-$ and $b \to s\nu\bar{\nu}$ 
transition decays by considering higher dimensional operators. We will consider only dimension six operators in our analysis. Moreover, it 
is believed that SMEFT analysis may be of great importance if no new particles are observed in LHC~\cite{Buchmuller:1985jz,Grzadkowski:2010es}.
 
The SMEFT Lagrangian corresponding to dimension six operators is expressed as~\cite{Grzadkowski:2010es}
\begin{equation}
 \mathcal{L}^{(6)}=\sum_i \frac{c_i}{\Lambda^2}\, \mathcal{Q}_i\,,
\end{equation}
where the relevant operators contributing to both $b \to s\, \nu\, \bar{\nu}$ and $b\,\to\,s\,\ell^+\,\ell^-$ decays are 
\begin{eqnarray}\label{smeft-o}
 \mathcal{Q}_{Hq}^{(1)}=i(\bar{q}_L \gamma_{\mu} q_L)H^\dagger D^{\mu} H, \hspace{0.5cm} 
 \mathcal{Q}_{Hq}^{(3)}=i(\bar{q}_L \gamma_{\mu} \tau^a q_L)H^\dagger D^{\mu} \tau_a H, \hspace{0.5cm}
 \mathcal{Q}_{Hd}=i(\bar{d}_R \gamma_{\mu} d_R)H^\dagger D^{\mu} H, \nonumber \\
 \mathcal{Q}_{ql}^{(1)}=(\bar{q}_L \gamma_{\mu} q_L)(\bar{l}_L \gamma^{\mu} l_L), \hspace{0.5cm}
 \mathcal{Q}_{ql}^{(3)}=(\bar{q}_L \gamma_{\mu} \tau^a q_L)(\bar{l}_L \gamma^{\mu} \tau_a l_L), \hspace{0.5cm}
 \mathcal{Q}_{dl}=(\bar{d}_R \gamma_{\mu} d_R)(\bar{l}_L \gamma^{\mu} l_L)\,
 \end{eqnarray}
and the operators contributing only to $b\,\to\,s\,\ell^+\,\ell^-$ decays are 
 \begin{eqnarray}\label{smeft-o2}
 \mathcal{Q}_{de}=(\bar{d}_R \gamma_{\mu} d_R)(\bar{e}_R \gamma^{\mu} e_R), \hspace{0.5cm}
 \mathcal{Q}_{qe}=(\bar{q}_L \gamma_{\mu} q_L)(\bar{e}_R \gamma^{\mu} e_R)\,.
\end{eqnarray}
At low energy, we can write down the most general $\Delta F=1$ effective Hamiltonian governing $b \to s\, \nu\, \bar{\nu}$ and 
$b\,\to\,s\,\ell^+\,\ell^-$ decays as~\cite{Altmannshofer:2009ma,Buras:2014fpa}
\begin{equation}
 \mathcal{H}_{eff}=-\frac{4 G_F}{\sqrt{2}}\, V_{tb} V_{ts}^{*}\, \frac{e^2}{16\pi^2}\, \sum_i C_i\, \mathcal{O}_i\, + h.c.,
\end{equation}
where $G_F$ is the Fermi coupling constant, $V_{tb}$ and $V_{ts}^{*}$ are the corresponding Cabibbo Kobayashi Maskawa (CKM) matrix elements. 
The operators corresponding to $b\,\to\,s\,\nu\,\bar{\nu}$ transition decays are represented by $\mathcal{O}_L$ and $\mathcal{O}_R$ with
WCs $C_L$ and $C_R$, respectively. The operators are
\begin{equation}
\mathcal{O}_L=(\bar{s}\gamma_{\mu}P_L b)(\bar{\nu}\gamma^{\mu}(1-\gamma_5)\nu), \hspace{0.5cm}
\mathcal{O}_R=(\bar{s}\gamma_{\mu}P_R b)(\bar{\nu}\gamma^{\mu}(1-\gamma_5)\nu)\,,
\end{equation}
where, $P_{L,R}=(1\mp\gamma_5)/2$ are the projection operators. In the SM, $C_R=0$ while $C_L = -6.38 \pm 0.06$. 
Similarly, the operators $\mathcal{O}_{9^{(\prime)},10^{(\prime)}}$ with corresponding WCs $C_{9^{(\prime)},10^{(\prime)}}$
contributing to $b\,\to\,s\,\ell^+\,\ell^-$ decays are represented by
\begin{equation}
 \mathcal{O}_9^{(\prime)}=(\bar{s}\gamma_{\mu}P_{L(R)} b)(\bar{l}\gamma^{\mu}l), \hspace{0.5cm}
\mathcal{O}_{10}^{(\prime)}=(\bar{s}\gamma_{\mu}P_{L(R)} b)(\bar{l}\gamma^{\mu}\gamma_5 l)\,,
\end{equation}
where the operators $\mathcal{O}_9^{\prime}$ and $\mathcal{O}_{10}^{\prime}$ exist purely in beyond the SM scenarios.
After electroweak symmetry breaking, the low energy SM WCs will get contribution from the dimension six operators of SMEFT. We write
$C_{9,10,L}$ and $C_{9^{\prime},10^{\prime},R}$ in terms of SMEFT WCs as~\cite{Buras:2014fpa}
 \begin{eqnarray}\label{ll2nu}
  C_9&=&C_9^{\rm SM}\, + \widetilde{c}_{qe}\, + \widetilde{c}_{ql}^{(1)}\, + \widetilde{c}_{ql}^{(3)}\, - \zeta \widetilde{c}_{Z} \nonumber \\
  C_{10}&=&C_{10}^{\rm SM}\, + \widetilde{c}_{qe}\, - \widetilde{c}_{ql}^{(1)}\, - \widetilde{c}_{ql}^{(3)}\, + \widetilde{c}_{Z} \nonumber \\
  {C}_{L}&=&C_{L}^{\rm SM}\, + \widetilde{c}_{ql}^{(1)}\, - \widetilde{c}_{ql}^{(3)}\, + \widetilde{c}_{Z} \nonumber \\
  C_{9}^{\prime}&=& \widetilde{c}_{de}\, + \widetilde{c}_{dl}\, - \zeta \widetilde{c}_{Z}^{\prime} \nonumber \\
  C_{10}^{\prime}&=& \widetilde{c}_{de}\, - \widetilde{c}_{dl}\, + \widetilde{c}_{Z}^{\prime} \nonumber \\
  {C}_{R}&=& \widetilde{c}_{dl}\, + \widetilde{c}_{Z}^{\prime}\,,
 \end{eqnarray}
where, $\widetilde{c}_{Z}=\frac{1}{2}(\widetilde{c}_{Hq}^{(1)}+\widetilde{c}_{Hq}^{(3)})$, $\widetilde{c}_{Z}^{\prime}=\frac{1}{2}
(\widetilde{c}_{Hd})$ and $\zeta \approx 0.08$ is the small vector coupling of $Z$ to charged leptons.
We refer to Refs.~\cite{Buras:2014fpa} for all the omitted details. 
Since we have two undetected neutrinos in the final state, we can only measure differential branching ratio as a function of $q^2$ for
$B \to P\,\nu\, \bar{\nu}$ decays, where $P$ stands for pseudoscalar meson. Whereas, we can measure differential branching ratio and 
longitudinal polarization fraction $F_L$ in case of $B \to V\,\nu\, \bar{\nu}$ decays, where $V$ stands for vector meson. 
All the expressions pertinent for our discussion are reported in Appendix~\ref{apdx2}.

\section{Results and discussions}\label{result}
\subsection{Input parameters}
For our numerical computation, we use several input parameters such as mass of mesons, quarks and leptons, CKM matrix element 
$|V_{tb} V_{ts}^{*}|$,
fine structure constant $\alpha$, Fermi coupling constant $G_F$ and the lifetime of parent $B_{(s)}$ meson.
For completeness, we report all the relevant input parameters taken from Ref.~\cite{ParticleDataGroup:2020ssz} in Table~\ref{tab_input}. 
Similarly, for $B\to K$ form factor inputs, we use the values obtained in LQCD~\cite{Bouchard:2013eph}. Again, for $B\to K^*$ and
$B_s \to \phi$ form factors, we use the combined LCSR and LQCD results as reported in Ref.~\cite{Bharucha:2015bzk}.
Moreover, we use the $B_s \to \eta^{(\prime)}$ form factor input parameters from Ref.~\cite{Duplancic:2015zna} that are obtained in the 
LCSR method. 

\begin{table}[htbp]
\centering
\setlength{\tabcolsep}{8pt} 
\renewcommand{\arraystretch}{1.5} 
\resizebox{\columnwidth}{!}{
\begin{tabular}{|c|c|c|c|c|c|c|c|c|c|}
\hline
Parameter & Value & Parameter & Value & Parameter & Value & Parameter & Value & Parameter & Value \\
\hline
\hline
$m_e$ & 0.000511 GeV &$m_{\mu}$ & 0.105658 GeV & $m_{B^+}$ & 5.27932 GeV & $m_{B^0}$ & 5.27963 GeV &$m_{B_s}$  & 5.3668 GeV\\
\hline
$m_{K^+}$ & 0.493677 GeV& $m_{K^{*0}}$ & 0.892 GeV& $m_\phi$ & 1.020 GeV& $m_{\eta}$ & 0.547862 GeV& $m_{\eta'}$ & 0.95778 GeV\\
\hline 
$m_b^{\bar{MS}}$ & 4.2 GeV& $m_c^{\rm \bar{MS}}$ & 1.28 GeV& $m_b^{pole}$ & 4.8 GeV& $\tau_{B^+}$ & $1.638 \times 10^{-12}$ s& $\tau _ {B^0}$ & $1.520 \times 10^{-12}$ s\\
\hline 
$\tau_{B_s}$ & $1.515 \times 10^{-12}$ s& $f_{B_s}$ & $0.225$ & $G_F$ & $1.1663787 \times 10^{-5}$ GeV$^{-2}$& $\alpha$ & 1/133.28 & $|V_{tb}V_{ts}^*|$ & 0.04088(55)\\
\hline 
\end{tabular}}
\caption{Theory input parameters}
\label{tab_input}
\end{table}

\subsection{Fit analysis of SMEFT coefficients}
Our main aim is to explore the consequences of $b \to s \ell \ell$ anomalies on several $b\,\to\,s\,\nu\,\bar{\nu}$ transition decays in a 
model independent SMEFT formalism. The SMEFT coefficients such as $\widetilde{c}_{ql}^{(1)}$, $\widetilde{c}_{ql}^{(3)}$ and 
$\widetilde{c}_{Z}$ corresponding to the left chiral currents appear in $C_{9,10}$ of $b \to s \ell \ell$ and $C_L$ of 
$b\,\to\,s\,\nu\,\bar{\nu}$ transitions.
Similarly, the SMEFT coefficients corresponding to the right chiral currents such as $\widetilde{c}_{dl}$ and $\widetilde{c}_{Z}^{\prime}$
appear in $C_{9,10}^\prime$ of $b \to s \ell \ell$ and $C_R$ of $b\,\to\,s\,\nu\,\bar{\nu}$ transitions, respectively. 
We consider several NP scenarios based on NP contributions from single
operators as well as from two different operators and try to find the scenario that best explains the anomalies present in $b \to s \ell \ell$
transition decays. To find the best fit values of these NP WCs, we perform a naive $\chi^2$ test with all the $b \to s \ell \ell$ experimental
data. The relevant $\chi^2$ is defined as
\begin{equation}
\chi^2= \sum_i  \frac{\Big ({\cal O}_i^{\rm th} -{\cal O}_i^{\rm exp} \Big )^2}{(\Delta {\cal O}_i^{\rm exp})^2+(\Delta {\cal O}_i^{\rm th})^2
}\,,
\end{equation}
where ${\cal O}_i^{\rm th}$ represents the theoretical value of each observable and ${\cal O}_i^{\rm exp}$ represents measured central value 
of the observables. $\Delta {\cal O}_i^{\rm th}$ and $\Delta {\cal O}_i^{\rm exp}$ represent the errors associated with the theory and 
experimental values, respectively. We perform two different fit analysis: Fit A and Fit B. In Fit A, we include a total of five measurements
for the evaluation of $\chi^2$, namely, $R_K$, $R_{K^*}$, $P_5^{\prime}$, $\mathcal{B}(B_s\, \to\, \phi\, \mu^+\,\mu^-)$ and 
$\mathcal{B}(B_s\, \to\, \mu^+\,\mu^-)$. In Fit B, we include only a subset of these five measurement for the evaluation of $\chi^2$, namely, 
$R_K$, $R_{K^*}$ and $\mathcal{B}(B_s\, \to\, \mu^+\,\mu^-)$. In Table~\ref{tab_fits1}, we report the best fit values of each SMEFT 
coefficients in several 1D and 2D scenarios for Fit A and Fit B. We also report the allowed $1\sigma$ range of each 1D coefficients.
In addition, we report the $\chi^2_{\rm min}$/d.o.f and the Pull$_{\rm SM} = \sqrt{\chi^2_{\rm SM}-\chi^2_{\rm NP}}$ for 
each scenarios. 
\begin{table}[ht]
\centering
\setlength{\tabcolsep}{8pt} 
\renewcommand{\arraystretch}{1.5} 
\begin{tabular}{|c||c|c||c|c||c|c|}
\hline
SMEFT couplings & \multicolumn{2}{c||}{Best fit} & \multicolumn{2}{c||}{$\chi^2_{\rm min}$/d.o.f} & \multicolumn{2}{c|}{Pull$_{\rm SM}$} \\
\cline{2-7}
 & Fit A & Fit B & Fit A & Fit B & Fit A & Fit B \\
\hline
\hline
\multirow{ 2}{*}{$\widetilde{c}_{ql}^{(1),(3)}$} & -0.667 & -0.460 & 4.095 & 0.533 & 2.48 & 2.58 \\ 
                                                          & (-1.196, -0.093) & (-0.899, -0.048) & & & &\\
\hline
\multirow{ 2}{*}{$\widetilde{c}_{Z}$} & 0.793 & 0.716 & 4.931 & 0.552 & 2.31 & 2.58 \\
                    & (0.155, 1.836) & (0.070, 1.461) & &  & &\\
\hline
$\widetilde{c}_{dl}$ & 0.025 & -0.076 & 10.593 & 6.953 & - & 0.49 \\
\hline
$\widetilde{c}_{Z}^{\prime}$ & -0.096 & 0.064 & 10.845 & 7.215 & - & - \\
\hline
\hline
$(\widetilde{c}_{ql}^{(1)},\widetilde{c}_{ql}^{(3)})$ & (-0.701, 0.103) & (-2.225, 1.759) & 4.113 & 0.537 & 2.48 & 2.58 \\
\hline
$(\widetilde{c}_{ql}^{(1),(3)},\widetilde{c}_{Z})$ & (-1.833, -1.849) & (-0.207, 0.396) & 3.695 & 0.493 & 2.56 & 2.59 \\
\hline
$(\widetilde{c}_{ql}^{(1),(3)},\widetilde{c}_{dl})$ & (-0.701, 0.103) & (-0.527, 0.169) & 3.878 & 0.132 & 2.53 & 2.66 \\
\hline
\bm{$(\widetilde{c}_{ql}^{(1),(3)},\widetilde{c}_{Z}^{\prime})$} & (-3.824, -4.905) & (-3.850, -4.994) & 0.324 & 0.047 & 3.15 & 2.67 \\
\hline
$(\widetilde{c}_{Z},\widetilde{c}_{dl})$ & (0.975, 0.038) & (0.764, 0.014) & 4.901 & 0.556 & 2.32 &  2.58 \\
\hline
\bm{$(\widetilde{c}_{Z},\widetilde{c}_{Z}^{\prime})$} & (4.560, -3.938) & (4.682, -3.985) & 1.040 & 0.086 & 3.04 & 2.67 \\
\hline
$(\widetilde{c}_{dl},\widetilde{c}_{Z}^{\prime})$ & (-0.596, -0.813) & (-0.779, -1.032) & 11.498 & 5.633 & - & 1.25 \\
\hline
\bm{$(\widetilde{c}_{ql}^{(1)}+\widetilde{c}_{ql}^{(3)},\widetilde{c}_{Z})$} & (-2.750, -2.293) & (4.099, 4.624) & 1.292 & 0.262 & 3.0 & 2.63 \\
\hline
$(\widetilde{c}_{ql}^{(1)}+\widetilde{c}_{ql}^{(3)},\widetilde{c}_{dl})$ & (-0.118, 0.933) & (2.252, 2.707) & 3.200 & 0.580 & 2.66 & 2.57 \\
\hline
\bm{$(\widetilde{c}_{ql}^{(1)}+\widetilde{c}_{ql}^{(3)},\widetilde{c}_{Z}^{\prime})$} & (-2.262, 1.519) & (4.484, -4.925) & 1.257 & 0.268 & 3.0 & 2.63 \\
\hline
\end{tabular}
\caption{Best fit values of SMEFT coefficients in several 1D and 2D scenarios. In Fit A, we include a total of five measurements
for the evaluation of $\chi^2$, namely, $R_K$, $R_{K^*}$, $P_5^{\prime}$, $\mathcal{B}(B_s\, \to\, \phi\, \mu^+\,\mu^-)$ and
$\mathcal{B}(B_s\, \to\, \mu^+\,\mu^-)$. In Fit B, we include only a subset of these five measurement for the evaluation of $\chi^2$, namely,
$R_K$, $R_{K^*}$ and $\mathcal{B}(B_s\, \to\, \mu^+\,\mu^-)$.}
\label{tab_fits1}
\end{table}
\begin{itemize}
 \item in Fit A, we have used five measured parameters for the the evaluation of $\chi^2$. Accordingly, the number of degrees of
freedom~(d.o.f) will be $5-1=4$ for $1D$ NP scenarios and $5-2 = 3$ for each $2D$ NP scenarios. To measure the disagreement of SM with the
data, we first obtain $\chi^2_{\rm min}$/d.o.f in the SM and it is found to be $10.264$.
The best fit value for each scenarios corresponds to the minimum $\chi^2_{\rm min}$ value. The allowed range of each 1D coefficients at $95\%$
confidence level~(CL) is obtained by imposing $\chi^2 \le 37.96$ constraint.
 \item In case of Fit B, we include only three measurement for the the evaluation of $\chi^2$. Accordingly, the number of d.o.f will be
$3-1=2$ for each $1D$ NP scenarios and $3-2=1$ for each $2D$ NP scenarios. In the SM, we have found $\chi^2_{\rm min}$/d.o.f to be $7.189$.
The allowed range of each 1D coefficients at $95\%$ CL is obtained by imposing $\chi^2 \le 11.98$ constraint.
\end{itemize}

From Table~\ref{tab_fits1}, it is clear that the coefficients $\widetilde{c}_{dl}$,
$\widetilde{c}_{Z}^{\prime}$ and $(\widetilde{c}_{dl}$, $\widetilde{c}_{Z}^{\prime})$ can not explain the anomalies present in
$b \to s \ell \ell$ data as the minimum $\chi^2$ values obtained for these scenarios are as large as or in some cases larger than that
of the SM $\chi^2$ value.  Hence we exclude them in the rest of our analysis. There, however, exists few $2D$ scenarios, namely,
$(\widetilde{c}_{ql}^{(1)},\widetilde{c}_{Z}^{\prime})$, $(\widetilde{c}_{ql}^{(3)},\widetilde{c}_{Z}^{\prime})$,
$(\widetilde{c}_{Z},\widetilde{c}_{Z}^{\prime})$, $(\widetilde{c}_{ql}^{(1)}+\widetilde{c}_{ql}^{(3)},\widetilde{c}_{Z})$ and
$(\widetilde{c}_{ql}^{(1)}+\widetilde{c}_{ql}^{(3)},\widetilde{c}_{Z}^{\prime})$ for which the Pull$_{\rm SM}$ is considerably larger
than the rest of the NP scenarios.
Moreover, these scenarios have better compatibility with $R_K$, $R_{K^*}$, $P_5^{\prime}$, $\mathcal{B}(B_s\, \to\, \phi\, \mu^+\,\mu^-)$,
$\mathcal{B}(B_s\, \to\, \mu^+\,\mu^-)$ experimental results. The compatibility of fit results with all $b \to s \ell \ell$ observables are
reported in Appendix~\ref{apdx1}.
Again, we do not find any special features in Fit B. For some $2D$ scenarios, we observe that Fit A serves as a better fit to the data
than Fit B. Hence, in all our future discussions, we will mainly focus on the Fit A results.
We now proceed to discuss the goodness of Fit A results with the measured values of $\mathcal B(B \to \, K^{(*)} \, \nu\,\bar{\nu})$.

\subsection{Additional constraints from $B \to \, K^{(*)} \, \nu\,\bar{\nu}$ decays}
We wish to determine the effect of the SMEFT coefficients on several $B \to \, K^{(*)} \, \nu\,\bar{\nu}$ decay observables, namely, 
$\mathcal B(B \to \, K^{(*)} \, \nu\,\bar{\nu})$, $F_L$, $\mathcal{R_{K^{(*)}}}$ and $\mathcal{R_{F_L}^{K^*}}$. In Table~\ref{tab_sm2}, we report
the central values and the corresponding $1\sigma$ uncertainty associated with each observable pertaining to 
$B \to \, K^{(*)} \, \nu\,\bar{\nu}$ decays in the SM and in the presence of several NP scenarios. To estimate the NP effects, we use
the best fit values of the 
SMEFT coefficients obtained in Fit A of Table.~\ref{tab_fits1}. 
In the SM, we obtain the branching fractions for both $B \to \, K^{(*)} \, \nu\,\bar{\nu}$ decays to be of $\mathcal{O}(10^{-6})$. Similarly, 
the ratios $\mathcal{R_K}$, $\mathcal{R_{K^*}}$ and $\mathcal{R_{F_L}^{K^*}}$ are found to be equal to $1$ in the SM. Hence any deviation 
from unity in these parameters could be a clear signal of beyond the SM physics.
There exist a few experiments that provide the upper bound of the branching ratio of $B \to \, K^{(*)} \, \nu\,\bar{\nu}$ decays.
At present, the upper bounds are found to be $\mathcal{B}({B \to\, K \, \nu\,\bar{\nu}}) < 11 \times 10^{-6}$ and 
$\mathcal{B}({B \to\, K^* \, \nu\,\bar{\nu}}) < 27 \times 10^{-6}$, respectively. Neglecting the theoretical uncertainty, we estimate  
the upper bound on $\mathcal{R_{K^{(*)}}}$ to be $\mathcal{R_K} \le 2.75$ and $\mathcal{R_{K^*}} \le 2.89$. Our observations are as follows.
\begin{itemize}
\item Values of $\mathcal B(B \to \, K^{(*)} \, \nu\,\bar{\nu})$ and  $\mathcal{R_{K^{(*)}}}$ obtained in each 1D NP scenarios with 
$\widetilde{c}_{ql}^{(1)}$, $\widetilde{c}_{ql}^{(3)}$ and $\widetilde{c}_{Z}$ SMEFT coefficients are compatible with 
experimental upper bound of $\mathcal{B}({B \to \, K^{(*)} \, \nu\,\bar{\nu}})$ and $\mathcal{R_{K^{(*)}}}$.

\item In case of 2D scenarios, we observe that the values of $\mathcal B(B \to \, K^{(*)} \, \nu\,\bar{\nu})$ and  $\mathcal{R_{K^{(*)}}}$ 
obtained with $(\widetilde{c}_{ql}^{(1)},\widetilde{c}_{Z}^{\prime})$ SMEFT coefficients are larger than the experimental upper bound. 
Although it can explain the anomalies present in $b \to s\,\ell^+\,\ell^-$ data, it, however, can not explain the $b \to s\nu\bar{\nu}$ data
simultaneously.

\item With $(\widetilde{c}_{ql}^{(1)},\widetilde{c}_{Z})$ and $(\widetilde{c}_{ql}^{(1)}+\widetilde{c}_{ql}^{(3)},\widetilde{c}_{Z})$ 
SMEFT coefficients, the value of $\mathcal B(B \to \, K^{(*)} \, \nu\,\bar{\nu})$ and  $\mathcal{R_{K^{(*)}}}$ are obtained to be quite large.
More precise data on $\mathcal B(B \to \, K^{(*)} \, \nu\,\bar{\nu})$ in future will put a severe constraint on these NP scenarios.

\item In the SM, $\mathcal{R_{F_L}^{K^*}} = 1$. Any deviation from unity is a clear signal of the presence of right handed currents.
It is evident from the Table~\ref{tab_sm2} that the value of $\mathcal{R_{F_L}^{K^*}}$ remains SM like for all the scenarios with left handed currents.
However, with the inclusion of right handed currents, its value seem to differ from unity. 
We see that the value of $F_L$ obtained in the presence of $(\widetilde{c}_{Z},\widetilde{c}_{Z}^{\prime})$ and 
$(\widetilde{c}_{ql}^{(3)},\widetilde{c}_{Z}^{\prime})$ coefficients are clearly distinguishable from SM prediction at more that $5\sigma$ 
level of significance.
\end{itemize}   
\begin{table}[ht!]
\centering
\setlength{\tabcolsep}{8pt} 
\renewcommand{\arraystretch}{1.5} 
\resizebox{\columnwidth}{!}{
\begin{tabular}{|c||c|c|c|c|c|c|}
\hline
SMEFT couplings & {$\mathcal{B}({B \to\, K \, \nu\,\bar{\nu}})\times 10^{-6}$} & {$\mathcal{R_K}$} & {$\mathcal{B}({B \to\, K^* \, \nu\,\bar{\nu}})\times 10^{-6}$} & {$\mathcal{R_{K^*}}$} & {$F_L ({B \to\, K^* \, \nu\,\bar{\nu}})$} & {$\mathcal{R_{F_L}^{K^*}}$} \\
\hline
\hline
SM                           & {$4.006 \pm 0.261$} & {1.000} & {$9.331 \pm 0.744$} & {1.000} & {$0.493 \pm 0.038$} & {1.000}  \\ 
\hline
{$\widetilde{c}_{ql}^{(1)}$} & $4.891 \pm 0.319$  & 1.221 & $11.394 \pm 0.908$ & 1.221 & $0.493 \pm 0.038$ & 1.000 \\ 
\hline
{$\widetilde{c}_{ql}^{(3)}$} & $3.209 \pm 0.209$ & 0.801 & $7.474 \pm 0.596$  & 0.801 & $0.493 \pm 0.038$ & 1.000 \\ 
\hline
{$\widetilde{c}_{Z}$} & $3.068 \pm 0.200$ & 0.766 & $7.147 \pm 0.570$ & 0.766 & $0.493 \pm 0.038$ & 1.000 \\
\hline
\hline
$(\widetilde{c}_{ql}^{(1)},\widetilde{c}_{ql}^{(3)})$ & $5.084 \pm 0.332$ & 1.269 & $11.843 \pm 0.944$ & 1.269 & $0.493 \pm 0.038$ & 1.000 \\
\hline
$(\widetilde{c}_{ql}^{(1)},\widetilde{c}_{Z})$ & $9.995 \pm 0.652$ & 2.495 & $23.284 \pm 1.856$ & 2.495 & $0.493 \pm 0.038$ & 1.000 \\
\hline
$(\widetilde{c}_{ql}^{(3)},\widetilde{c}_{Z})$ & $4.026 \pm 0.263$ & 1.005 & $9.378 \pm 0.748$ & 1.005 & $0.493 \pm 0.038$ & 1.000 \\
\hline
$(\widetilde{c}_{ql}^{(1)},\widetilde{c}_{dl})$ & $4.795 \pm 0.313$ & 1.197 & $11.732 \pm 0.941$ & 1.258 & $0.498 \pm 0.038$ & 1.009 \\
\hline
$(\widetilde{c}_{ql}^{(3)},\widetilde{c}_{dl})$ & $3.056 \pm 0.199$ & 0.763 & $7.568 \pm 0.608$ & 0.812 & $0.499 \pm 0.038$ & 1.011  \\
\hline
$(\widetilde{c}_{ql}^{(1)},\widetilde{c}_{Z}^{\prime})$ & $22.579 \pm 1.474$ & 5.637 & $14.035 \pm 0.963$ & 1.431 & $0.226 \pm 0.023$ & 0.481 \\
\hline
$(\widetilde{c}_{ql}^{(3)},\widetilde{c}_{Z}^{\prime})$ & $5.485 \pm 0.358$ & 1.369 & $3.197 \pm 0.223$ & 0.324 & $0.202 \pm 0.021$ & 0.432 \\
\hline
$(\widetilde{c}_{Z},\widetilde{c}_{dl})$ & $2.830 \pm 0.185$ & 0.706 & $6.750 \pm 0.540$ & 0.724 & $0.496 \pm 0.038$ & 1.004 \\
\hline
$(\widetilde{c}_{Z},\widetilde{c}_{Z}^{\prime})$ & $3.260 \pm 0.213$ & 0.814 & $2.141 \pm 0.146$ & 0.219 & $0.246 \pm 0.024$ & 0.521 \\
\hline
$(\widetilde{c}_{ql}^{(1)}+\widetilde{c}_{ql}^{(3)},\widetilde{c}_{Z})$ & $7.419 \pm 0.484$ & 1.852 & $17.284 \pm 1.378$ & 1.852 & $0.493 \pm 0.038$ & 1.000 \\
\hline
$(\widetilde{c}_{ql}^{(1)}+\widetilde{c}_{ql}^{(3)},\widetilde{c}_{dl})$ & $2.915 \pm 0.190$ & 0.728 & $11.370 \pm 0.958$ & 1.227 & $0.533 \pm 0.041$ & 1.072 \\
\hline
$(\widetilde{c}_{ql}^{(1)}+\widetilde{c}_{ql}^{(3)},\widetilde{c}_{Z}^{\prime})$ & $2.319 \pm 0.151$ & 0.579 & $12.857 \pm 1.111$ & 1.392 & $0.550 \pm 0.042$ & 1.103 \\
\hline
\end{tabular}}
\caption{Ratio of branching ratios $\mathcal{B}(B \to \, K^{(*)} \, \nu\,\bar{\nu})$, longitudinal polarization fraction of $K^*$ meson 
$F_{L}^{K^*}$ and the three ratios $\mathcal{R_K}$, $\mathcal{R_{K^*}}$ 
and $\mathcal{R_{F_L}^{K^*}}$ in the SM and with the best fit value of each SMEFT coefficients from Fit A analysis of Table.~\ref{tab_fits1}.}
\label{tab_sm2}
\end{table}

\begin{figure}[htbp]
\centering
\includegraphics[width=5.9cm,height=4.0cm]{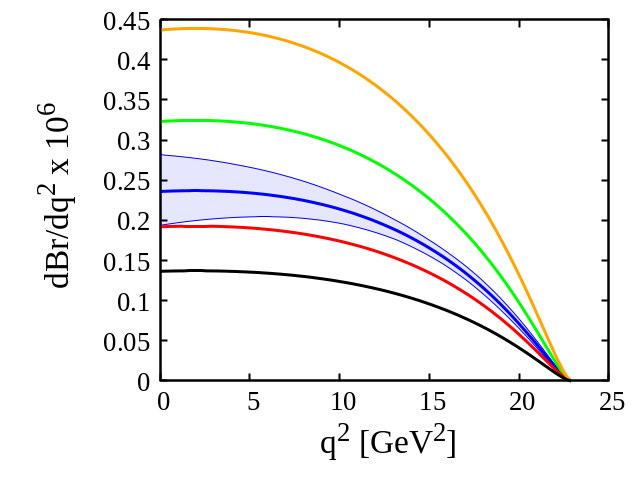}
\includegraphics[width=5.9cm,height=4.0cm]{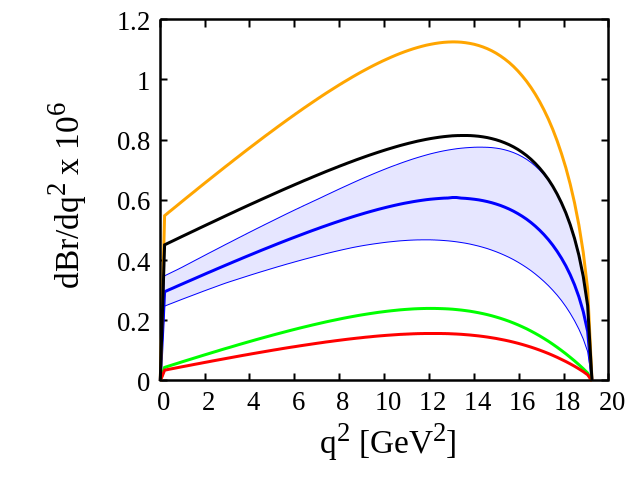}
\includegraphics[width=5.9cm,height=4.0cm]{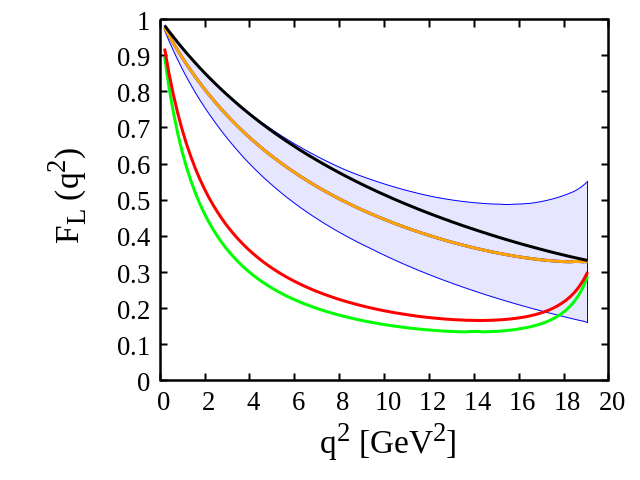}
\caption{$q^2$ dependence of differential branching ratios of $B \to \, K\, \nu\,\bar{\nu}$~(left) and 
$B \to \, K^{*} \,\nu\,\bar{\nu}$~(middle) decays and
longitudinal polarization fraction of $K^*$ meson $F_L^{K^*} (q^2)$~(right) in the SM and in few selected NP scenarios. 
The green, red, orange and black lines correspond to the best fit values of $(\widetilde{c}_{ql}^{(3)},\widetilde{c}_{Z}^{\prime})$,
$(\widetilde{c}_{Z},\widetilde{c}_{Z}^{\prime})$, $(\widetilde{c}_{ql}^{(1)}+\widetilde{c}_{ql}^{(3)},\widetilde{c}_{Z})$
and $(\widetilde{c}_{ql}^{(1)}+\widetilde{c}_{ql}^{(3)},\widetilde{c}_{Z}^{\prime})$, respectively. The corresponding SM central value
and the uncertainty band is shown with blue.}
\label{fig_con5}
\end{figure}
In Fig.~\ref{fig_con5}, we show the $q^2$ dependence of differential branching ratios and $K^*$ polarization fraction 
for the $B \to \, K^{(*)} \, \nu\,\bar{\nu}$ decays in the SM and for the best fit values of few selected new physics scenarios.
The SM central line and the corresponding $1\sigma$ uncertainty band is shown with blue color. 
The green, red, orange and black lines correspond to the best fit values of $(\widetilde{c}_{ql}^{(3)},\widetilde{c}_{Z}^{\prime})$, 
$(\widetilde{c}_{Z},\widetilde{c}_{Z}^{\prime})$, $(\widetilde{c}_{ql}^{(1)}+\widetilde{c}_{ql}^{(3)},\widetilde{c}_{Z})$
and $(\widetilde{c}_{ql}^{(1)}+\widetilde{c}_{ql}^{(3)},\widetilde{c}_{Z}^{\prime})$, respectively. We observe that the new physics 
contributions coming from these SMEFT coefficients are quite distinct.  
In case of $F_L(q^2)$, the contribution coming from 
$(\widetilde{c}_{ql}^{(3)},\widetilde{c}_{Z}^{\prime})$ and $(\widetilde{c}_{Z},\widetilde{c}_{Z}^{\prime})$ are more pronounced and they
are clearly distinguishable from SM contribution. In case differential branching ratio, the deviation from the SM prediction is maximum 
with $(\widetilde{c}_{ql}^{(1)}+\widetilde{c}_{ql}^{(3)},\widetilde{c}_{Z})$ NP scenario. The $K^*$ polarization fraction $F_L$
value, however, remains SM like as there is no right handed currents. 

We wish to quantify our results in terms of the independent parameters $\mathcal{R_K}$, $\mathcal{R_{K^*}}$ and $\mathcal{R_{F_L}^{K^*}}$. 
In the presence of $(\widetilde{c}_{ql}^{(3)},\widetilde{c}_{Z}^{\prime})$, value of $\mathcal{R_K}$ is increased by almost $\sim 30\%$ from 
the SM value, whereas, value of $\mathcal{R_{K^*}}$ and $\mathcal{R_{F_L}^{K^*}}$ are decreased by almost $\sim 70\%$ and $\sim 60\%$ 
from the SM prediction, respectively. 
In case of $(\widetilde{c}_{Z},\widetilde{c}_{Z}^{\prime})$,
we notice that the values of $\mathcal{R_K}$, $\mathcal{R_{K^*}}$ and $\mathcal{R_{F_L}^{K^*}}$ decreased by almost $\sim 20\%$, $\sim 80\%$ 
and $\sim 50\%$ from the SM predictions, respectively.
Similarly, with $(\widetilde{c}_{ql}^{(1)}+\widetilde{c}_{ql}^{(3)},\widetilde{c}_{Z})$, there is almost $\sim 80\%$ increment in 
$\mathcal{R_K}$ and $\mathcal{R_{K^*}}$, whereas, the value of
$\mathcal{R_{F_L}^{K^*}}$ remains SM like. This is because of the absence of right handed currents in this scenario. Finally, 
in case of $(\widetilde{c}_{ql}^{(1)}+\widetilde{c}_{ql}^{(3)},\widetilde{c}_{Z}^{\prime})$, value of $\mathcal{R_K}$ decreases by almost 
$\sim 40\%$, whereas, $\mathcal{R_{K^*}}$ and $\mathcal{R_{F_L}^{K^*}}$ increase by $\sim 40\%$ and $\sim 10\%$, respectively.

\subsection{Prediction of $B_s \to\, \eta \, \nu\,\bar{\nu}$, ${B_s \to\, \eta' \, \nu\,\bar{\nu}}$ and $B_s \to\, \phi\, \nu\,\bar{\nu}$ 
decay observables in SM and beyond}
Study of rare $B$ decays mediating via $b \to s\nu\bar{\nu}$ quark level transition is very well motivated as they can provide complimentary
information regarding NP in $b\,\to\,s\,\ell^+\,\ell^-$ quark level transition decays. To this end, we study several rare $B_s$ meson decays
such as $B_s \to\, \eta \, \nu\,\bar{\nu}$, ${B_s \to\, \eta' \, \nu\,\bar{\nu}}$ and $B_s \to\, \phi\, \nu\,\bar{\nu}$
proceeding via $b\,\to\,s\,\nu\,\bar{\nu}$ quark level transitions in a model independent SMEFT formalism. 
We give predictions of the branching fractions and $\phi$ polarization fraction in the SM and in the presence of several NP couplings.
For our NP analysis, we choose four NP scenarios, namely, $(\widetilde{c}_{ql}^{(3)},\widetilde{c}_{Z}^{\prime})$, $(\widetilde{c}_{Z},\widetilde{c}_{Z}^{\prime})$, $(\widetilde{c}_{ql}^{(1)}+\widetilde{c}_{ql}^{(3)},\widetilde{c}_{Z})$ and 
$(\widetilde{c}_{ql}^{(1)}+\widetilde{c}_{ql}^{(3)},\widetilde{c}_{Z}^{\prime})$, that provides the best solutions to the 
$b\,\to\,s\,\ell^+\,\ell^-$ anomalies.  
Interestingly, except $(\widetilde{c}_{ql}^{(1)}+\widetilde{c}_{ql}^{(3)},\widetilde{c}_{Z})$
the rest of the scenarios include the effects from right handed currents.  
In Table~\ref{tab_sm3}, we report the central values and the corresponding $1\sigma$ uncertainty associated with 
$\mathcal B(B_s \to\, (\eta \,,\eta'\,,\phi) \nu\,\bar{\nu})$ and $F_L(\phi)$ in the SM and in the presence of NP. 
We obtain the $1\sigma$ uncertainty associated with each of these observables by varying the input parameters such as the meson to meson
form factors and the CKM matrix elements within $1\sigma$ from their central values. 
In addition, we also quantify the results in terms of $\mathcal{R_\eta}$, $\mathcal{R_{\eta'}}$, $\mathcal{R_\phi}$ and 
$\mathcal{R_{F_L}^\phi}$. 

In the SM, we find the branching ratios of $B_s \to\, (\eta \,,\eta') \nu\,\bar{\nu}$ decays to be of 
$\mathcal{O}(10^{-6})$, whereas, for $B_s \to\ \phi\,\nu\,\bar{\nu}$ decays, it is found to be of $\mathcal{O}(10^{-5})$. The value of
$\phi$ polarization fraction is obtained to be $F_L = 0.537 \pm 0.030$. The 
NP effects can be easily quantified in terms of $\mathcal{R_{\eta\,,\eta'\,,\phi}}$ and $\mathcal{R_{F_L}^\phi}$.
We observe that $\mathcal B(B_s \to\, (\eta \,,\eta') \nu\,\bar{\nu})$ increases by almost $\sim 30\%$ in the presence of 
$(\widetilde{c}_{ql}^{(3)},\widetilde{c}_{Z}^{\prime})$, whereas, it decreases by almost $\sim 20\%$ due to the presence of 
$(\widetilde{c}_{Z},\widetilde{c}_{Z}^{\prime})$ NP couplings. Moreover, we observe a $\sim 80\%$ increment in the branching fraction due to 
$(\widetilde{c}_{ql}^{(1)}+\widetilde{c}_{ql}^{(3)},\widetilde{c}_{Z})$ NP coupling, whereas, with 
$(\widetilde{c}_{ql}^{(1)}+\widetilde{c}_{ql}^{(3)},\widetilde{c}_{Z}^{\prime})$ NP couplings, it decreases by almost $\sim 40\%$ 
with respect to the SM prediction.
In case of $B_s \to \phi\nu\bar{\nu}$ channel, we notice that $\mathcal B(B_s \to\, \phi\nu\,\bar{\nu})$ increases by almost $\sim 80\%$
with $(\widetilde{c}_{ql}^{(1)}+\widetilde{c}_{ql}^{(3)},\widetilde{c}_{Z})$ NP couplings. Similarly, we observe that the branching
fraction increases by almost $\sim 40\%$ in the presence of 
$(\widetilde{c}_{ql}^{(1)}+\widetilde{c}_{ql}^{(3)},\widetilde{c}_{Z}^{\prime})$, whereas, it decreases by almost $\sim 70\%-80\%$ 
with $(\widetilde{c}_{Z},\widetilde{c}_{Z}^{\prime})$ and
 $(\widetilde{c}_{ql}^{(3)},\widetilde{c}_{Z}^{\prime})$ NP couplings, respectively.
For $F_L$, we observe maximum deviation from the SM prediction with $(\widetilde{c}_{Z},\widetilde{c}_{Z}^{\prime})$ and 
$(\widetilde{c}_{ql}^{(3)},\widetilde{c}_{Z}^{\prime})$ NP couplings. 
Although, there is slight deviation observed due to $(\widetilde{c}_{ql}^{(1)}+\widetilde{c}_{ql}^{(3)},\widetilde{c}_{Z}^{\prime})$ NP
couplings, the deviation from the SM prediction, however, is quite small and it is not distinguishable from the SM.
\begin{table}[htbp]
\centering
\setlength{\tabcolsep}{8pt} 
\renewcommand{\arraystretch}{1.5} 
\resizebox{\columnwidth}{!}{
\begin{tabular}{|c||c|c|c|c|c|c|c|c|c|c|c|c|}
\hline
SMEFT couplings & $\mathcal{B}({B_s \to\, \eta \, \nu\,\bar{\nu}})\times 10^{-6}$ & $\mathcal{R_\eta}$ & $\mathcal{B}({B_s \to\, \eta' \, \nu\,\bar{\nu}})\times 10^{-6}$ & $\mathcal{R_{\eta'}}$ & $\mathcal{B}({B_s \to\, \phi\, \nu\,\bar{\nu}})\times 10^{-6}$ & $\mathcal{R_\phi}$ & $F_L ({B_s \to\, \phi\, \nu\,\bar{\nu}})$ & $\mathcal{R_{F_L}^\phi}$  \\
\hline
\hline
SM & $1.700 \pm 0.187$ & 1.000 & $1.673 \pm 0.232$ & 1.000 & $9.762 \pm 0.625$ & 1.000 & $0.537 \pm 0.030$ & 1.000 \\ 
\hline
$(\widetilde{c}_{ql}^{(3)},\widetilde{c}_{Z}^{\prime})$ & $2.327 \pm 0.256$ & 1.369 & $2.291 \pm 0.317$ & 1.369 & $3.007 \pm 0.128$ & 0.320 & $0.244 \pm 0.022$ & 0.437\\
\hline
$(\widetilde{c}_{Z},\widetilde{c}_{Z}^{\prime})$ & $1.383 \pm 0.152$ & 0.814 & $1.362 \pm 0.188$ & 0.814 & $2.047 \pm 0.091$ & 0.217 & $0.292 \pm 0.024$ & 0.526\\
\hline
$(\widetilde{c}_{ql}^{(1)}+\widetilde{c}_{ql}^{(3)},\widetilde{c}_{Z})$ & $3.148 \pm 0.347$ & 1.852 & $3.099 \pm 0.429$ & 1.852 & $18.083 \pm 1.157$ & 1.852 & $0.537 \pm 0.030$ & 1.000\\
\hline
$(\widetilde{c}_{ql}^{(1)}+\widetilde{c}_{ql}^{(3)},\widetilde{c}_{Z}^{\prime})$ & $0.984 \pm 0.108$ & 0.579 & $0.969 \pm 0.134$ & 0.579 & $13.714 \pm 0.950$ & 1.395 & $0.587 \pm 0.030$ & 1.101\\
\hline
\end{tabular}}
\caption{The branching ratios of $\mathcal{B}(B_s \to\, (\eta \,,\eta'\,,\phi) \nu\,\bar{\nu})$ and the longitudinal polarization fraction 
of the $\phi$ meson $F_{L}^{\phi}$ in the SM and with the best fit value of few selected 2D SMEFT scenarios of Fit A. The results are also 
quantified in terms of $\mathcal{R_\eta}$, $\mathcal{R_{\eta'}}$,$\mathcal{R_\phi}$ and $\mathcal{R_{F_L}^\phi}$.}
\label{tab_sm3}
\end{table}

\begin{figure}[htbp]
\centering
\includegraphics[width=5.9cm,height=4.0cm]{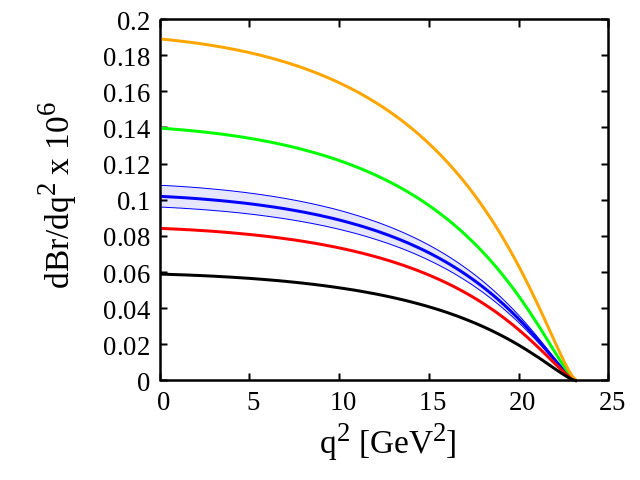}
\includegraphics[width=5.9cm,height=4.0cm]{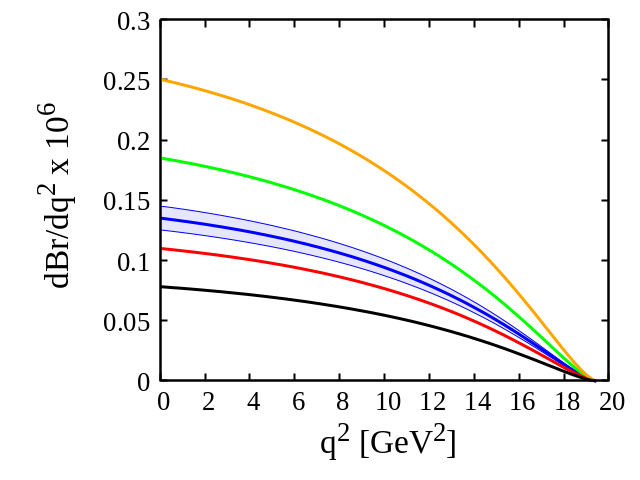}

\includegraphics[width=5.9cm,height=4.0cm]{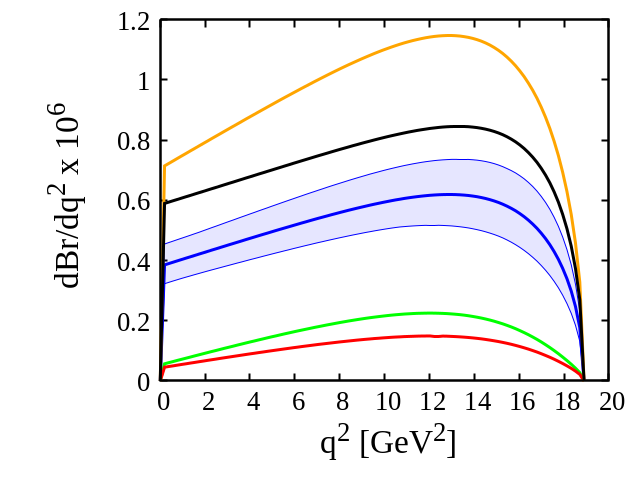}
\includegraphics[width=5.9cm,height=4.0cm]{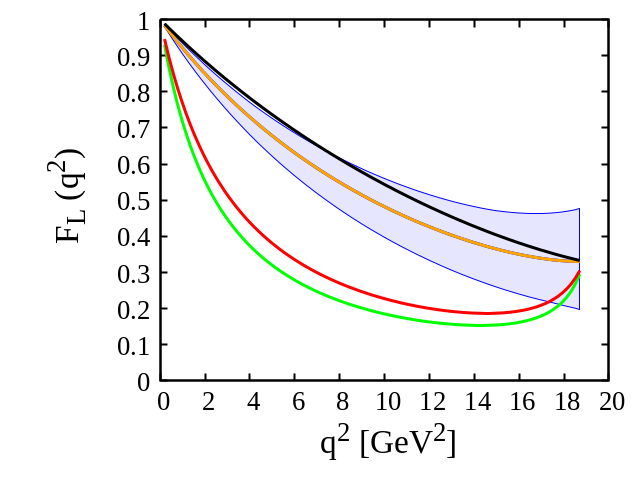}
\caption{We show, in the top panel, the $q^2$ dependence of differential branching fractions of $B_s \to\, \eta\, \nu\,\bar{\nu}$~(left) and 
$B_s \to\, \eta'\, \nu\,\bar{\nu}$~(right) decays. 
In the bottom panel, we show the differential branching fraction~(left) and the $\phi$ polarization fraction~(right) of 
$B_s \to\, \phi\, \nu\,\bar{\nu}$ decay, respectively. 
The SM central line and the corresponding error band is shown with blue. 
The green, red, orange and black lines correspond to the best fit values of $(\widetilde{c}_{ql}^{(3)},\widetilde{c}_{Z}^{\prime})$,
$(\widetilde{c}_{Z},\widetilde{c}_{Z}^{\prime})$, $(\widetilde{c}_{ql}^{(1)}+\widetilde{c}_{ql}^{(3)},\widetilde{c}_{Z})$
and $(\widetilde{c}_{ql}^{(1)}+\widetilde{c}_{ql}^{(3)},\widetilde{c}_{Z}^{\prime})$, respectively.}
\label{fig_con4}
\end{figure}
In Fig.~\ref{fig_con4}, we display the $q^2$ dependence of differential branching ratios
and $\phi$ polarization fraction $F_L(q^2)$ for $B_s \to\, (\eta \,,\eta'\,,\phi) \nu\,\bar{\nu}$ decays in the SM and in the presence of
NP couplings. The SM central line and the corresponding uncertainty band obtained at $95\%$ CL are represented with blue
color. The green, red, orange and black lines correspond to the best fit values of $(\widetilde{c}_{ql}^{(3)},\widetilde{c}_{Z}^{\prime})$,
$(\widetilde{c}_{Z},\widetilde{c}_{Z}^{\prime})$, $(\widetilde{c}_{ql}^{(1)}+\widetilde{c}_{ql}^{(3)},\widetilde{c}_{Z})$ and 
$(\widetilde{c}_{ql}^{(1)}+\widetilde{c}_{ql}^{(3)},\widetilde{c}_{Z}^{\prime})$ from Fit A of Table.~\ref{tab_fits1}, respectively. Our observations 
are as follows.

\begin{itemize}
\item The differential branching ratio for  $B_s \to\, (\eta\,,\eta')\,\nu\,\bar{\nu}$ decays is enhanced at all $q^2$ for 
$(\widetilde{c}_{ql}^{(3)},\widetilde{c}_{Z}^{\prime})$ and $(\widetilde{c}_{ql}^{(1)}+\widetilde{c}_{ql}^{(3)},\widetilde{c}_{Z})$,
whereas, it is reduced at all $q^2$ for $(\widetilde{c}_{Z},\widetilde{c}_{Z}^{\prime})$ and 
$(\widetilde{c}_{ql}^{(1)}+\widetilde{c}_{ql}^{(3)},\widetilde{c}_{Z}^{\prime})$. All the NP scenarios are distinguishable from the SM 
prediction at more than $3\sigma$ and they are quite distinct from each other. The deviation from the SM prediction is more pronounced in case 
of $(\widetilde{c}_{ql}^{(1)}+\widetilde{c}_{ql}^{(3)},\widetilde{c}_{Z})$ NP scenario.
 
\item The differential branching ratio for $B_s \to \phi\nu\bar{\nu}$ decays is enhanced at all $q^2$ for 
$(\widetilde{c}_{ql}^{(1)}+\widetilde{c}_{ql}^{(3)},\widetilde{c}_{Z})$ and
$(\widetilde{c}_{ql}^{(1)}+\widetilde{c}_{ql}^{(3)},\widetilde{c}_{Z}^{\prime})$, whereas, it is reduced at all $q^2$ in case of
 $(\widetilde{c}_{ql}^{(3)},\widetilde{c}_{Z}^{\prime})$ and $(\widetilde{c}_{Z},\widetilde{c}_{Z}^{\prime})$ NP scenarios. 
All the NP scenarios are distinguishable from the SM prediction at more than $3\sigma$. The deviation observed is more pronounced in case
of $(\widetilde{c}_{ql}^{(3)},\widetilde{c}_{Z}^{\prime})$,
$(\widetilde{c}_{Z},\widetilde{c}_{Z}^{\prime})$ and $(\widetilde{c}_{ql}^{(1)}+\widetilde{c}_{ql}^{(3)},\widetilde{c}_{Z})$ NP scenarios.
 
\item The $\phi$ polarization fraction $F_L(q^2)$ for the $B_s \to\, \phi\, \nu\,\bar{\nu}$ decay is distinct from SM only in the presence of 
$(\widetilde{c}_{Z},\widetilde{c}_{Z}^{\prime})$, $(\widetilde{c}_{ql}^{(3)},\widetilde{c}_{Z}^{\prime})$ and 
$(\widetilde{c}_{ql}^{(1)}+\widetilde{c}_{ql}^{(3)},\widetilde{c}_{Z}^{\prime})$ that includes the contribution from right handed currents.
In case of $(\widetilde{c}_{ql}^{(1)}+\widetilde{c}_{ql}^{(3)},\widetilde{c}_{Z})$, it is SM like. 
The deviation from the SM prediction observed with $(\widetilde{c}_{Z},\widetilde{c}_{Z}^{\prime})$ and 
$(\widetilde{c}_{ql}^{(3)},\widetilde{c}_{Z}^{\prime})$
is quite significant and they are distinguishable from the SM prediction at more than $5\sigma$. 
A slight deviation is observed with $(\widetilde{c}_{ql}^{(1)}+\widetilde{c}_{ql}^{(3)},\widetilde{c}_{Z}^{\prime})$
and it is not distinguishable from the SM prediction.
\end{itemize}

\section{Conclusion}\label{concl}
Motivated by the long standing anomalies in $B$ decays with charged leptons in the final state undergoing 
$b\, \to\, s\, \mu^+\, \mu^-$ quark level transition, we study several $B$ meson decays, namely, $B \, \to \, K^{(*)} \,\nu \, \bar{\nu}$, 
$B_s \to\, (\eta,\eta') \, \nu\,\bar{\nu}$ and $B_s \, \to \, \phi \, \nu \, \bar{\nu}$ mediating via $b \, \to \,s\, \nu \, \bar{\nu}$ 
quark level transition. 
Our primarily goal of this study is intended to analyze the consequences of latest $b\, \to\, s\, \mu^+\, \mu^-$ anomalies on 
$b \, \to \,s\, \nu \, \bar{\nu}$ decays in a model independent approach. We use the standard model effective field theory formalism 
constructed out of new operators of dimension six corresponding to the arbitrary Wilson coefficients.
We study several decay observables pertaining to these decay modes in the SM and in the presence of various SMEFT coefficients in several 
1D and 2D scenarios. 
We perform a naive $\chi^2$ fit to the $b\, \to\, s\, \mu^+\, \mu^-$ data, namely, 
$R_K$, $R_{K^*}$, $P_5^{\prime}$, $\mathcal{B}(B_s\, \to\, \phi\, \mu^+\,\mu^-)$ and $\mathcal{B}(B_s\, \to\, \mu^+\,\mu^-)$, to find the best
fit values of all the SMEFT coefficients in several 1D and 2D scenarios. 
We observe that the pull corresponding to 2D scenarios are comparatively better than the 1D scenarios. 
In particular, the fit results of $(\widetilde{c}_{ql}^{(1),(3)},\widetilde{c}_{Z}^{\prime})$, 
$(\widetilde{c}_{Z},\widetilde{c}_{Z}^{\prime})$, $(\widetilde{c}_{ql}^{(1)}+\widetilde{c}_{ql}^{(3)},\widetilde{c}_{Z})$ and 
$(\widetilde{c}_{ql}^{(1)}+\widetilde{c}_{ql}^{(3)},\widetilde{c}_{Z}^{\prime})$ of 2D SMEFT scenarios
show better compatibility with all the five $b\,\to\,s\,\ell \ell$ measured data. 
We also check the goodness of the fit results with the additional constraints coming from the experimental upper bounds of
$\mathcal B(B \to \, K^{(*)} \, \nu\,\bar{\nu})$. We observe that, although, $(\widetilde{c}_{ql}^{(1)},\widetilde{c}_{Z}^{\prime})$ provides
a better solution to the $b\,\to\,s\,\ell \ell$ data, it, however, cannot explain the existing $b \to s\nu\bar{\nu}$ data. 
The estimated value of $\mathcal B(B \to \, K^{(*)} \, \nu\,\bar{\nu})$ with the best fit value of 
$(\widetilde{c}_{ql}^{(1)},\widetilde{c}_{Z}^{\prime})$ exceeds the experimental upper bounds of 
$\mathcal B(B \to \, K^{(*)} \, \nu\,\bar{\nu})$. 

In case of branching ratio, we observe a significant deviation from the SM prediction in all the four NP scenarios. 
All the NP scenarios are distinguishable from the SM prediction at more than $3\sigma$ significance. The deviation observed with
$(\widetilde{c}_{ql}^{(3)},\widetilde{c}_{Z}^{\prime})$,
$(\widetilde{c}_{Z},\widetilde{c}_{Z}^{\prime})$ and $(\widetilde{c}_{ql}^{(1)}+\widetilde{c}_{ql}^{(3)},\widetilde{c}_{Z})$ are
more pronounced. Similarly, for $F_L$, The deviation from the SM prediction observed with $(\widetilde{c}_{Z},\widetilde{c}_{Z}^{\prime})$ 
and $(\widetilde{c}_{ql}^{(3)},\widetilde{c}_{Z}^{\prime})$
is quite significant and they are distinguishable from the SM prediction at more than $5\sigma$.
Study of $B_s \to\, (\eta,\eta') \, \nu\,\bar{\nu}$ and $B_s \, \to \, \phi \, \nu \, \bar{\nu}$ decay modes 
are very well motivated theoretically as well as experimentally as they can, in principle, provide complementary information 
regarding NP in $b\,\to\,s\,\ell^+ \ell^-$ decays. 
Experimental investigations of decay observables in $b \, \to \,s\, \nu \, \bar{\nu}$ in the future will definitely help us in identifying 
the possible new physics Lorentz structures in $b\,\to\,s\,\ell^+ \ell^-$ decays.
In particular, measurement of $F_L$ will be very crucial to not only examine the effects of right handed currents but also to 
distinguish between various new physics models.

\acknowledgments 
NR would like to thank CSIR for the financial support in this work.
 
\appendix

\section{Differential decay distribution for $B \to (P,V)\, \nu\, \bar{\nu}$ decays}
\label{apdx2}
The differential decay distribution for $B \to P\,\nu\,\bar{\nu}$ decays, where $P$ denote
pseudoscalar meson, can be written as~\cite{Altmannshofer:2009ma,Buras:2014fpa}
\begin{equation}
 \frac{d\Gamma(B \to P\, \nu\, \bar{\nu})}{dq^2}=\frac{G_{F}^{2}\alpha^2}{256\pi^5 m_B^3}|V_{tb}\, V_{ts}^{*}|^{2}\, \lambda^{3/2}(m_B^2, m_P^2, q^2)\,[f_+ (q^2)]^2 \, |{C}_{L}+{C}_{R}|^2\,.
\end{equation}
Similarly, for $B \to V\, \nu\, \bar{\nu}$, it can be written as
\begin{equation}
 \frac{d\Gamma(B \to V\, \nu\, \bar{\nu})}{dq^2}=3\, \left[ |A_{\perp}|^2 + |A_{\parallel}|^2 +|A_0|^2 \right]\,,
\end{equation}
where, $|A_{\perp}|, |A_{\parallel}|, |A_0|$ are the $B\to V$ transversity amplitudes which can be expressed in terms of form factors and 
Wilson coefficients as
\begin{eqnarray}
 A_{\perp}(q^2)=\frac{2N\,\sqrt{2\lambda(m_B^2, m_V^2, q^2)}}{m_B^2}\,[{C}_{L}+{C}_{R}]\,\frac{V(q^2)}{\left[1+\frac{m_V}{m_B}\right]}\, ; \hspace{1cm} \nonumber
 A_{\parallel}(q^2)=-2N\sqrt{2}\,\left[1+\frac{m_V}{m_B}\right]\,[{C}_{L}-{C}_{R}]\,A_1 (q^2)\, ; \nonumber
 \end{eqnarray}
 \begin{equation}
 A_0 (q^2)=-\frac{N\,[{C}_{L}-{C}_{R}]\,m_B^2}{m_V\, \sqrt{q^2}}\,
 \left( \left[ 1- \frac{m_V^2}{m_B^2} - \frac{q^2}{m_B^2}\right] \,\left[1+\frac{m_V}{m_B}\right]\, A_1 (q^2)\, - \, \frac{\lambda(m_B^2, m_V^2, q^2)}{m_B^2} \frac{A_2 (q^2)}{\left[1+\frac{m_V}{m_B}\right]} \right)
\end{equation}
with
\begin{equation}
 N= |V_{tb}\, V_{ts}^{*}|\, \left[ \frac{G_{F}^2\, \alpha^2\,q^2\,\sqrt{\lambda(m_B^2, m_V^2, q^2)}}{3\times 2^{10}\pi^5\, m_B} \right]^{1/2}
\,.
\end{equation}
Here $N$ is the normalization factor and $q^2$ is the invariant mass of the neutrino-antineutrino pair.
The factor $\lambda$ is defined as $\lambda(a,b,c)=a^2+b^2+c^2-2(ab+bc+ca)$.
The $B\to P$ and $B\to V$ form factors are defined in terms of $f_+ (q^2)$, $V(q^2)$, $A_1 (q^2)$, $A_2 (q^2)$, respectively.
Similarly, longitudinal polarization fraction of the final vector meson can be written as
\begin{equation}
 F_{L}=\frac{3\,|A_0|^2}{d\Gamma/dq^2}\,.
\end{equation}

In addition to the differential branching ratio and polarization fraction, we define 
$\mathcal{R_P}$, $\mathcal{R_V}$ and $\mathcal{R_{F_L}^V}$ where, $\mathcal{P}$ and $\mathcal{V}$ represent pseudoscalar and vector mesons,
respectively. They are expressed in terms of the three real parameters $\epsilon$, $\eta$ and $\kappa_\eta$ as~\cite{Buras:2014fpa}.
\begin{equation}
 \mathcal{R_P}=(1-2\eta)\epsilon^2, \hspace{1cm}
 \mathcal{R_V}=(1+\kappa_\eta \eta)\epsilon^2, \hspace{1cm}
 \mathcal{R_{F_L}^V}=\frac{1+2\eta}{1+\kappa_\eta \eta}
\end{equation}
where,
\begin{equation}
 \epsilon =\frac{\sqrt{|C_L|^2 + |C_R|^2}}{|C_L^{\rm SM}|}, \hspace{1cm}
 \eta =\frac{- \rm Re(C_L C_R^*)}{|C_L|^2 + |C_R|^2}
\end{equation}
\begin{equation}
 \kappa_\eta= 2\frac{\int dq^2 \left(\rho_{A_1} (q^2)+\rho_{A_{12}} (q^2)-\rho_V (q^2)\right)}{\int dq^2 \left(\rho_{A_1} (q^2)+\rho_{A_{12}} (q^2)+\rho_V (q^2)\right)}
\end{equation}
where,
\begin{equation}
 \rho_P (q^2) = \frac{\lambda^{3/2}(m_B^2, m_P^2, q^2)}{m_B^4}\, [f_+ (q^2)]^2, \hspace{1cm} \nonumber 
 \rho_V (q^2) = \frac{2q^2 \lambda^{3/2}(m_B^2, m_V^2, q^2)}{(m_B + m_V)^2 m_B^4}\, [V(q^2)]^2, \nonumber
 \end{equation}
 \begin{equation}
 \rho_{A_1} (q^2) = \frac{2q^2 \lambda^{1/2}(m_B^2, m_V^2, q^2)(m_B + m_V)^2}{m_B^4}\, [A_1(q^2)]^2, \hspace{1cm} \nonumber
 \rho_{A_{12}} (q^2) = \frac{64 m_V^2 \lambda^{1/2}(m_B^2, m_V^2, q^2)}{m_B^2}\, [A_{12}(q^2)]^2\,,
\end{equation}
where, $\rho_i$ is rescaled form factors. 
It is important to note that the value of $\mathcal{R_P}$ is independent of decay mode as it only depends on the WCs $C_{L,R}$. 
However, $\mathcal{R_V}$ and $\mathcal{R_{F_L}^V}$ do depend on the decay mode through the factor $\kappa_\eta$.
The contribution from $\kappa_\eta$ is observed to be very tiny for $B \to \, K^{*} \, \nu\,\bar{\nu}$ and 
$B_s \to \, \phi \, \nu\,\bar{\nu}$ decays.

\section{Best estimates of $R_K$, $R_{K^*}$, $P_5^{\prime}$, $\mathcal{B}(B_s\, \to\, \phi\, \mu^+\,\mu^-)$ and 
$\mathcal{B}(B_s\, \to\, \mu^+\,\mu^-)$ in the presence of several 1D and 2D SMEFT coefficients from Fit A and Fit B analysis.}
\label{apdx1}

\begin{table}[htbp]
\centering
\setlength{\tabcolsep}{8pt} 
\renewcommand{\arraystretch}{1.5} 
\resizebox{\columnwidth}{!}{
\begin{tabular}{|c|c|c|c|c|c||c|c|c|}
\hline
SMEFT & \multicolumn{5}{c||}{Fit-A} & \multicolumn{3}{c|}{Fit-B} \\ \cline{2-9}
 & $R_K$ & $R_{K^*}$ & $P'_5$ & $\mathcal{B}(B_s\to \phi \mu \mu)\times 10^{-7}$ & $\mathcal{B}(B_s \to \mu \mu)\times 10^{-9}$ & $R_K$ & $R_{K^*}$ & $\mathcal{B}(B_s \to \mu \mu)\times 10^{-9}$ \\
\hline
\hline
Expt. values$\to$ & $0.846 \pm 0.060$ & $0.685 \pm 0.150$ & $-0.21 \pm 0.15$ & $1.44 \pm 0.21$ & $3.09 \pm 0.484$ & $0.846 \pm 0.060$ & $0.685 \pm 0.150$ & $3.09 \pm 0.484$ \\
$1\sigma \to$ & (0.786, 0.906) & (0.535, 0.835) & (-0.36, -0.06) & (1.23, 1.65) & (2.606, 3.574) & (0.786, 0.906) & (0.535, 0.835) & (2.606, 3.574) \\ 
$2\sigma \to$ & (0.726, 0.966) & (0.385, 0.985) & (-0.51, 0.09) & (1.02, 1.86) & (2.122, 4.085) & (0.726, 0.966) & (0.385, 0.985) & (2.122, 4.085) \\
\hline
\hline
\multirow{ 2}{*}{$\widetilde{c}_{ql}^{(1),(3)}$} & 0.730 & 0.724 & -0.661 & 2.148 & 2.807 & 0.814 & 0.802 & 3.120 \\ 
                           & (0.543, 0.963) & (0.545, 0.958) & (-0.952, -0.617) & (1.597, 3.225) & (2.147, 3.796) & (0.647, 0.986) & (0.639, 0.979) & (2.521, 3.906) \\
\hline
\multirow{ 2}{*}{$\widetilde{c}_{Z}$} & 0.823 & 0.797 & -0.741 & 2.353 & 2.649 & 0.839 & 0.812 & 2.887 \\
                    & (0.635, 0.968) & (0.589, 0.956) & (-1.025, -0.693) & (1.739, 3.195) & (1.436, 3.657) & (0.697, 0.987) & (0.651, 0.981) & (1.850, 3.848) \\
\hline
\hline
\multirow{ 2}{*}{$(\widetilde{c}_{ql}^{(1)},\widetilde{c}_{ql}^{(3)})$} & 0.756 & 0.752 & -0.668 & 2.256 & 2.916 & 0.813 & 0.801 & 3.122 \\
                                                      & (0.544, 0.967) & (0.548, 0.958) & (-0.949, -0.613) & (1.609, 3.207) & (2.150, 3.752) & (0.646, 0.985) & (0.640, 0.979) & (2.513, 3.902) \\
\hline
\multirow{ 2}{*}{$(\widetilde{c}_{ql}^{(1),(3)},\widetilde{c}_{Z})$} & 0.719 & 0.768 & -0.449 & 2.244 & 3.851 & 0.824 & 0.807 & 3.047 \\
                                               & (0.485, 0.972) & (0.541, 0.972) & (-0.993, 0.276) & (1.635, 3.206) & (0.500, 5.267) & (0.640, 1.008) & (0.644, 0.979) & (1.435, 4.592) \\
\hline
\multirow{ 2}{*}{$(\widetilde{c}_{ql}^{(1),(3)},\widetilde{c}_{dl})$} & 0.756 & 0.681 & -0.695 & 2.044 & 2.630 & 0.851 & 0.725 & 2.893 \\
                                                & (0.546, 1.047) & (0.395, 0.991) & (-1.056, -0.620) & (1.224, 3.094) & (1.416, 3.888) & (0.643, 1.047) & (0.446, 1.136) & (1.409, 4.176) \\
\hline
\multirow{ 2}{*}{\bm{$(\widetilde{c}_{ql}^{(1),(3)},\widetilde{c}_{Z}^{\prime})$}} & 0.833 & 0.611 & -0.199 & 1.764 & 2.957 & 0.850 & 0.645 & 3.077 \\
                                                        & (0.482, 1.165) & (0.286, 0.952) & (-1.035, 0.519) & (0.831, 3.121) & (0.813, 4.581) & (0.640, 1.039) & (0.401, 1.013) & (1.528, 4.289) \\
\hline
\multirow{ 2}{*}{$(\widetilde{c}_{Z},\widetilde{c}_{dl})$} & 0.804 & 0.743 & -0.757 & 2.212 & 2.376 & 0.838 & 0.799 & 2.797 \\
                                         & (0.599, 1.052) & (0.535, 0.978) & (-1.058, -0.574) & (1.546, 3.172) & (1.292, 3.671) & (0.655, 1.043) & (0.551, 1.061) & (1.475, 4.045) \\
\hline
\multirow{ 2}{*}{\bm{$(\widetilde{c}_{Z},\widetilde{c}_{Z}^{\prime})$}} & 0.862 & 0.690 & -0.209 & 2.043 & 2.569 & 0.844 & 0.723 & 2.947 \\
                                                 & (0.560, 1.176) & (0.422, 0.981) & (-1.039, 0.546) & (1.243, 3.348) & (0.602, 4.598) & (0.648, 1.050) & (0.540, 1.039) & (1.435, 4.383) \\
\hline
\multirow{ 2}{*}{\bm{$(\widetilde{c}_{ql}^{(1)}+\widetilde{c}_{ql}^{(3)},\widetilde{c}_{Z})$}} & 0.783 & 0.894 & -0.231 & 1.861 & 3.251 & 0.838 & 0.779 & 3.125 \\
                                                                        & (0.488, 1.107) & (0.54, 1.355) & (-0.984, 0.553) & (0.532, 3.076) & (0.567, 5.087) & (0.644, 1.010) & (0.597, 1.137) & (1.452, 4.683) \\
\hline
\multirow{ 2}{*}{$(\widetilde{c}_{ql}^{(1)}+\widetilde{c}_{ql}^{(3)},\widetilde{c}_{dl})$} & 0.789 & 0.790 & -0.645 & 1.898 & 2.397 & 0.818 & 0.812 & 3.150  \\
                                                                         & (0.551, 1.086) & (0.546, 1.072) & (-0.924, -0.114) & (0.491, 3.134) & (0.510, 3.889) & (0.640, 0.998) & (0.631, 1.071) & (1.524, 4.693) \\
\hline 
\multirow{ 2}{*}{\bm{$(\widetilde{c}_{ql}^{(1)}+\widetilde{c}_{ql}^{(3)},\widetilde{c}_{Z}^{\prime})$}} & 0.809 & 0.886 & -0.389 & 1.717 & 2.763 & 0.822 & 0.763 & 3.173 \\
                                                                                 & (0.494, 1.103) & (0.561, 1.374) & (-1.013, 0.563) & (0.564, 3.126) & (0.497, 5.194) & (0.640, 1.009) & (0.596, 1.136) & (1.436, 4.709) \\                                                                   
\hline 
\end{tabular}}
\caption{Values of $R_K$, $R_{K^*}$, $P_5^{\prime}$, $\mathcal{B}(B_s\, \to\, \phi\, \mu^+\,\mu^-)$ and 
$\mathcal{B}(B_s\, \to\, \mu^+\,\mu^-)$ with the best fit value of each SMEFT coefficients from Fit A and Fit B analysis of 
Table.~\ref{tab_fits1}. 
In the first row we report the experimental central value and the corresponding $1\sigma$ and $2\sigma$ range of each of these observables.} 
\label{b2sll}
\end{table}

\FloatBarrier

\end{document}